\documentclass[twocolumn,showpacs,preprintnumbers,amsmath,amssymb]{revtex4}

\usepackage{graphicx}
\usepackage{dcolumn}
\usepackage{bm}

\newcommand{\epsfigbox}[5]{%
\begin{figure} \vspace{#3}%
\includegraphics[width=6.0cm]{#2}%
\caption{ \label{fig:#1} #5} \vspace{#4}
\end{figure}}

\newcommand{\epsfigpox}[5]{%
\begin{figure} \vspace{#3}%
\includegraphics[width=4.5cm]{#2}%
\caption{ \label{fig:#1} #5} \vspace{#4}
\end{figure}}

\newcommand{\epsfigdox}[5]{%
\begin{figure} \vspace{#3}%
\includegraphics[width=8.0cm]{#2}%
\caption{ \label{fig:#1} #5} \vspace{#4}
\end{figure}}

\begin{document}

\title{Shell model method for Gamow-Teller Transitions in heavy, deformed nuclei}

\author{
Zao-Chun Gao$^{1,2,3}$, Yang Sun$^{1}$, Y.-S. Chen$^{3,4}$}

\affiliation{$^{1}$Joint Institute for Nuclear Astrophysics and
Department of Physics,\\ University of Notre Dame, Notre Dame,
Indiana 46556, USA \\
$^{2}$Department of Physics and Astronomy and National
Superconducting Cyclotron Laboratory, Michigan State University,
East Lansing, Michigan 48824, USA \\
$^{3}$China Institute of Atomic Energy, P.O. Box
275(18), Beijing 102413, P.R. China \\
$^{4}$Institute of Theoretical Physics, Academia Sinica, Beijing
100080, P.R. China }

\begin{abstract}
A method for calculation of Gamow-Teller transition rates is
developed by using the concept of the Projected Shell Model (PSM).
The shell model basis is constructed by superimposing
angular-momentum-projected multi-quasiparticle configurations, and
nuclear wave functions are obtained by digonalizing the two-body
interactions in these projected states. Calculation of transition
matrix elements in the PSM framework is discussed in detail, and
the effects caused by the Gamow-Teller residual forces and by
configuration-mixing are studied. With this method, it may become
possible to perform a state-by-state calculation for $\beta$-decay
and electron-capture rates in heavy, deformed nuclei at finite
temperatures. Our first example indicates that, while
experimentally known Gamow-Teller transition rates from the ground
state of the parent nucleus are reproduced, stronger transitions
from some low-lying excited states are predicted to occur, which
may considerably enhance the total decay rates once these nuclei
are exposed to hot stellar environments.
\end{abstract}

\date{\today}
\maketitle

\section{Introduction}

The knowledge on weak interaction processes is one of the most
important ingredients for resolving astrophysical problems. The
first systematical work on stellar weak-interaction rates was
performed by Fuller, Fowler, and Newman
\cite{FFN1,FFN2,FFN3,FFN4}, who recognized the decisive role
played by the Gamow-Teller (GT) transitions. Due to their
pioneering work, the study of stellar weak interaction rates
requested by astrophysics becomes essentially a nuclear structure
problem, in which the actual decay rates are determined by the
microscopic inside of nuclear many-body systems. It has been
suggested that the nuclear shell model, i.e. a full
diagonalization of an effective Hamiltonian in a chosen model
space, is the most preferable method for GT transition
calculations. This was noticed early by Aufderheide {\it et al.}
\cite{ABRM93}, and has recently been emphasized by Langanke and
Mart\'inez-Pinedo \cite{LM03}.

For a theoretical model employed in GT transition calculations, it
is generally required that the model can reproduce a wide range of
structure properties of relevant nuclei. It has been shown that
the state-of-the-art shell-model diagonalization method is indeed
capable of performing such calculations. For example, Wildenthal
and Brown \cite{Wild84,BW85} obtained nuclear wave functions in
the full $sd$-shell model space, which were successfully applied
to calculation of GT rates in the $sd$ shell nuclei \cite{Oda94}.
By using the method developed by the Strasbourg-Madrid group
\cite{SM94}, Langanke and Mart\'inez-Pinedo \cite{LM01} made the
shell-model GT rates available also for the $pf$ shell nuclei.
Still, these sophisticated calculations are tractable only for
nuclei up to the mass-60 region, and cannot be applied to heavier
nuclei which play important roles in the nuclear processes in
massive stars.

In the long history of the nuclear shell-model development,
tremendous effort has been devoted to extending the shell-model
capacity from its traditional territory to heavier shells. Despite
the great progress made in recent years, it seems impossible to
treat an arbitrarily large nuclear system in a spherical shell
model framework due to the unavoidable problem of dimension
explosion. One is thus compelled to seek judicious schemes to deal
with large nuclear systems. The central issue has been the
shell-model truncation. There are many different ways of
truncating a shell-model space. While in principle, it does not
matter how to prepare a model basis, it is crucial in practice to
use the most efficient one. In this regard, we recognize the fact
that except for a few lying in the vicinity of shell closures,
most nuclei in the nuclear chart are deformed. This naturally
suggests for a shell model calculation to use a {\it deformed}
basis to incorporate the physics in large systems. That is the
philosophy that the Projected Shell Model (PSM) \cite{PSM} and the
several important generalizations
\cite{Sheikh99,Chen01,Sun02,Sun03,Gao06} are based on.

The present article reports on the new development for calculation
of GT transition rates in the PSM framework. Before a detailed
description of the work, we mention a few attractive features in
our approach, which may be relevant for future astrophysical
applications.

\begin{itemize}

\item The PSM utilizes single particle bases generated by deformed
mean-field models yet carries out a shell-model diagonalization
like the conventional shell model. Conceptually, the PSM bridges
two important nuclear structure methods: the deformed mean-field
approach and the conventional shell model, and takes the
advantages of both. On the one hand, as a shell model, the PSM can
be applied to any heavy, deformed nuclei without a size
limitation. On the other hand, unlike the mean-field models or
models with an average nature, the PSM wave functions contain
correlations beyond mean-field and the states are written in the
laboratory frame having definite quantum numbers such as
angular-momentum and parity. These are needed properties when the
wave functions are employed in transition calculations.

\item Because of the way the PSM constructs its basis, the
dimension of the model space is small (usually in the range of
$10^2 - 10^4$). With this size of basis, a state-by-state
evaluation of GT transition rates is computationally feasible.
This feature is important because in stellar environments with
finite temperatures, the usual situation is that the thermal
population of excited states in a parent nucleus sets up
connections to many states in a daughter by the GT operator.
However, our current knowledge on GT transitions from excited
nuclear states is very poor, and in many cases, it must rely on
theoretical calculations.

\item The calculation of forbidden transitions involves nuclear
transitions between different harmonic oscillator shells and thus
requires multi-shell model spaces. Such a calculation is not
feasible for most of the conventional shell models working in
one-major shell bases. The PSM is a multi-shell shell model. This
feature is desired particularly when forbidden transitions are
dominated.

\item Isomeric states belong to a special group of nuclear states
because of their long half-lives. The existence of isomeric states
in nuclei could alter significantly the elemental abundances
produced in nucleosynthesis. There are cases in which an isomer of
sufficiently long lifetime can change the paths of reactions
taking place and lead to a different set of elemental abundances
\cite{Sun05a,Sun05b}. It has been shown that the PSM is indeed
capable of describing the detailed structure of isomeric states
\cite{Sun04}.

\end{itemize}

The paper is organized as follows. In Sec. II, we briefly
introduce the PSM concept and describe how shell model
diagonalization is carried out in the angular-momentum-projected
bases. Technical details of calculation of transition matrix
elements in the projected bases are given in Sec. III. Our first
example of a GT transition calculation is illustrated in Sec. IV,
where we first validate the model by comparing the calculated
structure properties (energy levels and electromagnetic
transitions) with experiment. The obtained GT transition results
are then discussed and the effects caused by the Gamow-Teller
residual forces and by configuration mixing are studied. Finally,
the work is summarized and an outlook on future applications is
given in Sec. V.

\section{Outline of the model}

The Projected Shell Model \cite{PSM} works with the following
scheme. It begins with the deformed Nilsson single particle basis,
with pairing correlations incorporated into the basis by a BCS
calculation for the Nilsson states. The Nilsson-BCS calculation
defines a deformed quasiparticle (qp) basis. Then the
angular-momentum (and if necessary, also particle-number, parity)
projection is performed on the qp basis to form a shell model
space in the laboratory frame. Finally a two-body Hamiltonian is
diagonalized in this projected space.

Let $\left|\Phi\right>$ be the qp vacuum and $a^\dagger_{\nu}$ and
$a^\dagger_{\pi}$ the qp creation operators, with the index $\nu$
($\pi$) denoting the neutron (proton) quantum numbers and running
over selected single-qp states for each configuration. The
multi-qp configurations of the PSM are given for the four kinds of
nuclei as follows:
\begin{widetext}
\begin{equation}
\begin{array}{rl}
\mbox{e-e:} & \{ \left|\Phi \right\rangle, a^\dagger_{\nu_i}
a^\dagger_{\nu_j} \left|\Phi \right\rangle, a^\dagger_{\pi_i}
a^\dagger_{\pi_j} \left|\Phi \right\rangle, a^\dagger_{\nu_i}
a^\dagger_{\nu_j} a^\dagger_{\pi_k} a^\dagger_{\pi_l} \left|\Phi
\right\rangle, \cdots \},
\vspace{4pt}\\
\mbox{odd-$\nu$:} & \{ a^\dagger_{\nu_i} \left|\Phi \right\rangle,
a^\dagger_{\nu_i} a^\dagger_{\nu_j} a^\dagger_{\nu_k} \left|\Phi
\right\rangle, a^\dagger_{\nu_i} a^\dagger_{\pi_j}
a^\dagger_{\pi_k} \left|\Phi \right\rangle, a^\dagger_{\nu_i}
a^\dagger_{\nu_j} a^\dagger_{\nu_k} a^\dagger_{\pi_l}
a^\dagger_{\pi_m} \left|\Phi \right\rangle, \cdots \},
\vspace{4pt}\\
\mbox{odd-$\pi$:} & \{ a^\dagger_{\pi_i} \left|\Phi \right\rangle,
a^\dagger_{\pi_i} a^\dagger_{\nu_j} a^\dagger_{\nu_k} \left|\Phi
\right\rangle, a^\dagger_{\pi_i} a^\dagger_{\pi_j}
a^\dagger_{\pi_k} \left|\Phi \right\rangle, a^\dagger_{\pi_i}
a^\dagger_{\pi_j} a^\dagger_{\pi_k} a^\dagger_{\nu_l}
a^\dagger_{\nu_m} \left|\Phi \right\rangle, \cdots \},
\vspace{4pt}\\
\mbox{o-o:} & \{ a^\dagger_{\nu_i} a^\dagger_{\pi_j} \left|\Phi
\right\rangle, a^\dagger_{\nu_i} a^\dagger_{\nu_j}
a^\dagger_{\nu_k} a^\dagger_{\pi_l} \left|\Phi \right\rangle,
a^\dagger_{\nu_i} a^\dagger_{\pi_j} a^\dagger_{\pi_k}
a^\dagger_{\pi_l} \left|\Phi \right\rangle, a^\dagger_{\nu_i}
a^\dagger_{\nu_j} a^\dagger_{\nu_k} a^\dagger_{\pi_l}
a^\dagger_{\pi_m} a^\dagger_{\pi_n} \left|\Phi \right\rangle,
\cdots \}. \label{qpset}
\end{array}
\end{equation}
\end{widetext}
Note that the PSM works with multiple harmonic-oscillator shells
for both neutrons and protons. The indices $\nu$ and $\pi$ in
(\ref{qpset}) are general; for example, a 2-qp state can be of
positive parity if both quasiparticles $i$ and $j$ are from the
major $N$-shells that differ in $N$ by $\Delta N = 0, 2, \dots$,
or of negative parity if $i$ and $j$ are from those $N$-shells
that differ by $\Delta N = 1, 3, \dots$. In bases (\ref{qpset}),
``$\cdots$" denotes those configurations having a higher order of
quasiparticles, which may practically be ignored in the present
study. On the other hand, the bases (\ref{qpset}) can be easily
enlarged if necessary. If the configurations denoted by
``$\cdots$" are completely included, one recovers the full shell
model space written in the representation of qp excitation.

The shell model basis states can then be constructed by the
projection technique. Without losing generality, the PSM
wave-function can be written as
\begin{equation}
\left|\Psi^{\sigma}_{IM}\right> = \sum _{K \kappa}
f^{\sigma}_{IK_\kappa}\,\hat P^I_{MK}\left|\Phi_\kappa \right> ,
\label{wf}
\end{equation}
where $\left|\Phi_\kappa\right\rangle$ denotes the qp-basis given
in (1), and
\begin{equation}
\hat{P}^{I}_{MK} = \frac{2I+1}{8\pi^2}\int d\Omega
D^{I}_{MK}(\Omega)\hat{R}(\Omega) \label{PN}
\end{equation}
is the angular momentum projection operator \cite{RS80}. In
Eq.\,(\ref{PN}), $D^{I}_{MK}$ is the $D$-function \cite{Ed60},
$\hat{R}$ the rotation operator, and $\Omega$ the solid angle. If
one keeps the axial symmetry in the deformed basis, $D^{I}_{MK}$
in Eq.\,(\ref{PN}) reduces to the small $d$-function and the three
dimensions in $\Omega$ reduce to one. The energies and wave
functions (given in terms of the coefficients
$f^{\sigma}_{IK_\kappa}$ in Eq. (\ref{wf})) are obtained by
solving the following eigen-value equation:
\begin{equation}
\sum_{K'\kappa'}(H_{K\kappa, K' \kappa'}^I-E^\sigma_IN_{K\kappa,
K' \kappa'}^I)f^\sigma_{IK_{\kappa'}}=0
\end{equation}
where $H^I_{K\kappa, K'\kappa'}$ and $N^I_{K\kappa, K'\kappa'}$
are respectively the matrix elements of the Hamiltonian and the
norm
\begin{equation}
\begin{array}{rcl}
H_{K\kappa, K' \kappa'}^I &=& \langle \Phi_\kappa|\hat H \hat
P^I_{KK'}|\Phi_{\kappa'}\rangle
\vspace{6pt}\\
N_{K\kappa, K' \kappa'}^I &=& \langle \Phi_\kappa|\hat
P^I_{KK'}|\Phi_{\kappa'}\rangle .
\end{array}
\end{equation}

The PSM uses a large single-particle space, which ensures that the
collective motion and the cross-shell interplay are defined
microscopically by accommodating a sufficiently large number of
active nucleons. It usually includes three (four) major
harmonic-oscillator shells each for neutrons and protons in a
calculation for deformed (superdeformed or superheavy) nuclei.
However, the shell model dimension in Eq. (\ref{wf}) is small.
This means that each of the configurations in (\ref{wf}) is a
complex combination of spherical shell model basis states.
Although the dimension where the final diagonalization is carried
out is small, it is huge in terms of original shell model
configurations. In this sense, the PSM is truly a shell model in a
truncated multi-major-shell space.

The Hamiltonian in the present study consists of the separable
forces
\begin{equation}
\hat H = \hat H_0 + \hat H_{QP} + \hat H_{GT},
\label{Hamilt}
\end{equation}
which represent different kinds of characteristic correlations
between valence particles. It has the single-particle term $\hat
H_0$, the quadrupole+pairing force $\hat H_{QP}$ with inclusion of
the quadrupole-pairing term, and the Gamow-Teller force of the
charge-exchange terms $\hat H_{GT}$. $\hat H_0$ contains a set of
properly adjusted single-particle energies in the Nilsson scheme
\cite{Nilsson69}. The second force, $\hat H_{QP}$, contains three
terms \cite{PSM}
\begin{equation}
\hat H_{QP} = - \frac12 \chi_{QQ} \sum_\mu \hat Q^\dagger_{2\mu}
\hat Q^{}_{2\mu} - G_M \hat P^\dagger \hat P - G_Q \sum_\mu \hat
P^\dagger_{2\mu} \hat P^{}_{2\mu} ,
\label{QP}
\end{equation}
which are quadrupole-quadrupole, monopole-pairing, and
quadrupole-pairing interactions, respectively. The strength of the
quadrupole-quadrupole force $\chi_{QQ}$ is determined in a
self-consistent manner that it would give the empirical
deformation as predicted in mean-field calculations \cite{PSM}.
The monopole-pairing strength is taken to be the form
\begin{equation}
G_M = {{G_1 \mp G_2{{N-Z}\over A}} \over A},
\label{Mpairing}
\end{equation}
where ``+" (``$-$") is for protons (neutrons), $N$, $Z$, and $A$
are respectively the neutron number, proton number, and mass
number, and $G_1$ and $G_2$ are the coupling constants adjusted to
yield the known odd-even mass differences. The quadrupole-pairing
strength $G_Q$ is taken to be about 20$\%$ of $G_M$, as is often
assumed in the PSM calculations \cite{PSM}. It has been shown in
many previous publications that such a set of interaction can
reasonably well describe structures in heavy nuclei. The one-body
operators (for each kind of nucleons) in Eq.\,(\ref{QP}) are of
the standard form
\begin{eqnarray}
\hat Q_{2\mu}&=&\sum_{\alpha,\alpha'}\
\langle\alpha|\hat{Q}_{2\mu}|\alpha'\rangle\ c^\dagger_\alpha
c_{\alpha'}
\vspace{4pt}\nonumber\\
\hat P^\dagger &=& \frac12\sum_\alpha  c^\dagger_\alpha
c^\dagger_{\bar{\alpha}}
\vspace{4pt}\nonumber\\
\hat P^\dagger_{2\mu}&=&  \frac12 \sum_{\alpha,\alpha'}\
\langle\alpha|\hat{Q}_{2\mu}|\alpha'\rangle c^\dagger_\alpha
c^\dagger_{\bar{\alpha}'} \nonumber
\end{eqnarray}
where $c^\dagger_\alpha$ is the nucleon creation operator, with
$\alpha$ standing for the quantum numbers of a single-particle
state in the spherical basis ($\alpha\equiv\{ n\ell jm\}$). The
time reversal of $c_\alpha$ is defined as $c_{\bar{\alpha}}\equiv
(-)^{j-m}c_{n \ell j-m}$.

The last force, $\hat H_{GT}$ in Eq. (\ref{Hamilt}), is the
Gamow-Teller force
\begin{eqnarray}
\hat H_{GT} &=& +\ 2\chi_{GT} \sum_\mu \hat \beta^-_{1\mu}(-1)^\mu
\hat \beta^+_{1-\mu} \nonumber\\
&& -\ 2\kappa_{GT} \sum_\mu \hat \Gamma^-_{1\mu}(-1)^\mu \hat
\Gamma^+_{1-\mu}.
\label{GT}
\end{eqnarray}
This is a charge-dependent separable interaction with both
particle-hole (ph) and particle-particle (pp) channels, which act
between protons and neutrons. This type of force has been used by
several authors \cite{Soloviev88,Homma96,Civitarese98,Moreno06} in
the study of single- and double-$\beta$ decay. The ph and pp
interactions are defined to be repulsive and attractive,
respectively, when the strength parameters $\chi_{GT}$ and
$\kappa_{GT}$ take positive values. The pp interaction, which was
introduced by Kuz'min and Soloviev \cite{Soloviev88}, is a
neutron-proton pairing force in the $J^\pi=1^+$ channel. In the
present calculation, we adopt the interaction strengths
\begin{equation}
\begin{array}{rcl}
\chi_{GT}&=&{23 / A}
\vspace{4pt}\\
\kappa_{GT}&=&{7.5 / A},
\end{array}
\label{GTstrength}
\end{equation}
from the original work of Kuz'min and Soloviev \cite{Soloviev88}.
We notice that there are different versions for these parameters,
for example, the one extracted from systematical studies by Homma
{\it et al.} \cite{Homma96}. The one-body operators appearing in
Eq. (\ref{GT}) are defined as
\begin{equation}
\begin{array}{rcl}
\hat
\beta^-_{1\mu}&=&\sum_{\pi,\nu}\langle\pi|\sigma_\mu\tau_-|\nu\rangle
c^\dagger_{\pi} c_{\nu}, ~~~~~ \hat \beta^+_{1\mu} =
(-)^\mu(\beta^-_{1-\mu})^\dagger
\vspace{6pt}\\
\hat
\Gamma^-_{1\mu}&=&\sum_{\pi,\nu}\langle\pi|\sigma_\mu\tau_-|\nu\rangle
c^\dagger_{\pi} c^\dagger_{\bar{\nu}}, ~~~~~ \hat \Gamma^+_{1\mu}
= (-)^\mu(\Gamma^-_{1-\mu})^\dagger ,
\end{array}
\label{bgOperator}
\end{equation}
where $\sigma$ and $\tau$ are the Pauli spin operator and the
isospin operator, respectively. As one will see later in an
example of the $^{164}$Ho $\rightarrow$ $^{164}$Dy transition, the
Gamow-Teller force is important for a correct reproduction of the
GT strengths.

We finally mention that the Hamiltonian assumed in the form of Eq.
(\ref{Hamilt}) may need to be extended when specific quantities or
transition processes are studied. For example, the spin-dipole
force would be necessary to reproduce first-forbidden transitions.

\section{Transition matrix elements in the projected states}

The probability of Gamow-Teller transition from an initial state
of spin $I_i$ to a final state $I_f$ is defined as
\begin{eqnarray}
B(GT,I_i\rightarrow I_f)=\frac{2I_f+1}{2I_i+1}|\langle
\Psi_{I_f}\|\hat \beta^{\pm}\|\Psi_{I_i} \rangle|^2 ,
\label{BGT}
\end{eqnarray}
where the operators $\hat \beta^{\pm}$ are those given in Eq.
(\ref{bgOperator}). A GT transition involves a transfer of one
unit of angular momentum, which is described by the Pauli spin
operators $\sigma_\mu$. The isospin operators $\tau_\pm$ transform
a neutron into a proton, and {\it vice versa}.

The heart of the present development is the evaluation of
transition matrix element in the angular-momentum-projected bases.
Let us start with the PSM wave function in Eq. (\ref{wf})
\begin{equation}
\left|\Psi^\sigma_{IM}\right> = \sum_{K\kappa}
f^\sigma_{IK_\kappa} \hat P^I_{MK} \left|\Phi_\kappa \right> .
\nonumber
\end{equation}
In a general weak interaction process with isospin exchange, the
initial and final states must correspond to different nuclear
systems. Therefore, two different sets of quasiparticle are
generally involved. For example, in a calculation of $\beta$-decay
from an odd-odd parent to an even-even daughter nucleus,
$\left|\Phi_\kappa\right>$ for the initial odd-odd system has the
form
\begin{widetext}
\begin{equation}
\left|\Phi_\kappa(a)\right> \equiv \{ a^\dagger_\nu a^\dagger_\pi
\left|a\right>, a^\dagger_\nu a^\dagger_\nu a^\dagger_\nu
a^\dagger_\pi \left|a\right>, a^\dagger_\nu a^\dagger_\pi
a^\dagger_\pi a^\dagger_\pi \left|a\right>, a^\dagger_\nu
a^\dagger_\nu a^\dagger_\nu a^\dagger_\pi a^\dagger_\pi
a^\dagger_\pi \left|a\right>, \cdots \},
\end{equation}
and for the final even-even system,
\begin{equation}
\left|\Phi_\kappa(b)\right>
\equiv \{ \left|b\right>, b^\dagger_\nu b^\dagger_\nu
\left|b\right>, b^\dagger_\pi b^\dagger_\pi \left|b\right>,
b^\dagger_\nu b^\dagger_\nu b^\dagger_\pi b^\dagger_\pi
\left|b\right>, \cdots \}.
\end{equation}
Here, $\{a^\dagger\}$ and $\{b^\dagger\}$ are two different sets
of quasiparticle operators associated with the quasiparticle vacua
$\left|a\right>$ and $\left|b\right>$, respectively.

To compute the matrix element of a tensor operator of rank
$\lambda$ in the projected states, the matrix element of the
operator $\hat P^{I_f}_{K_f M_f} \hat T_{\lambda\mu} \hat
P^{I_i}_{M_i K_i}$ has to be evaluated. A useful expression was
derived in Ref. \cite{PSM}
\begin{equation}
\hat P^{I_f}_{K_f M_f} \hat T_{\lambda\mu} \hat P^{I_i}_{M_i K_i}=
\langle I_i M_i \lambda \mu|I_f M_f\rangle \sum_\nu \langle
I_iK_f-\nu,\lambda\nu|I_fK_f\rangle \hat T_{\lambda\nu} \hat
P^{I_i}_{K_f-\nu K_i} ,
\end{equation}
which follows from the transformation property of a tensor
operator of rank $\lambda$ under rotation as well as the reduction
theorem of a product of two $D$-functions \cite{Ed60}. With the
help of this relation, the reduced matrix element in Eq.
(\ref{BGT}) can be written as
\begin{equation}
\langle \Psi_{I_f}\|\hat \beta^{\pm}\|\Psi_{I_i}\rangle =
\sum_{K_i\kappa,K_f\kappa'} f_{I_iK_i\kappa}f_{I_fK_f\kappa'}
\left\{\sum_\mu\langle I_iK_f-\mu,1\mu|I_fK_f\rangle
\langle\Phi_{\kappa'}(b) |\hat \beta^{\pm}_{1\mu} \hat
P^{I_i}_{K_f-\mu K_i}|\Phi_{\kappa}(a)\rangle\right\} .
\label{ME1}
\end{equation}
\end{widetext}

If we assume an axial symmetry in nuclei, as for the example we
shall discuss in the next section, the general three-dimensional
angular momentum projection is reduced to a problem of
one-dimensional projection, with the projector having the
following form
\begin{equation}
\hat P^I_{MK} = \left(I+{1\over 2}\right) \int^\pi_0 d\beta ~~{\rm
sin} \beta ~~d^I_{MK}(\beta) ~~\hat R_y(\beta) \label{proj}
\end{equation}
with
\begin{equation}
\hat R_y(\beta) = e^{-i\beta \hat J_y}.
\end{equation}
In Eq. (\ref{proj}), $d^I_{MK}(\beta)$ is the small-$d$ function
and $\beta$ is one of the Euler angels \cite{Ed60}. Inserting the
projector (\ref{proj}) into $\langle\Phi_{\kappa'}(b)|\hat
\beta^{\pm}_{1\mu} \hat P^{I_i}_{K_f-\mu
K_i}|\Phi_{\kappa}(a)\rangle$ in Eq. (\ref{ME1}), one obtains
\begin{widetext}
\begin{equation}
\left<\Phi_{\kappa'}(b)\right|\hat \beta^{\pm}_{1\mu} \hat
P^{I_i}_{K_f-\mu K_i} \left|\Phi_{\kappa}(a)\right> =
\left(I_i+{1\over 2}\right) \int^\pi_0 d\beta ~~{\rm sin} \beta
~~d^{I_i}_{K_f-\mu K_i}(\beta) \left<\Phi_{\kappa'}(b)\right| \hat
\beta^{\pm}_{1\mu} \hat R_y(\beta) \left|\Phi_{\kappa}(a)\right>.
\end{equation}
\end{widetext}
The problem is now reduced to evaluation of the matrix element
\begin{equation}
\left<\Phi_{\kappa'}(b)\right| \hat \beta^{\pm}_{1\mu} \hat
R_y(\beta) \left|\Phi_{\kappa}(a)\right>,
\label{RMT}
\end{equation}
which is the problem of calculating the $\hat \beta$ operator
sandwiched by a multi-qp state $\left| \Phi_{\kappa'}(b)\right>$
and a {\it rotated} multi-qp state $\hat R_y(\beta) \left|
\Phi_{\kappa}(a)\right>$, with $a$ and $b$ characterizing
different qp sets.

To calculate $\left<\Phi_{\kappa'}(b)\right| \hat
\beta^{\pm}_{1\mu} \hat R_y(\beta) \left|\Phi_{\kappa}(a)\right>$,
one must compute the following types of contractions for the
Fermion operators
\begin{eqnarray}
A_{ij} &=& \left<b\right| [\beta] a^\dagger_ia^\dagger_j
 \left|a\right> = [V(\beta)U^{-1}(\beta)]_{ij},
\nonumber\\
B_{ij} &=& \left<b\right|b_ib_j [\beta] \left|a\right> =
 [U^{-1}(\beta)V(\beta)]_{ij},
\label{ABCmatrix}\\
C_{ij} &=& \left<b\right|b_i [\beta] a^\dagger_j \left|a\right> =
[U^{-1}(\beta)]_{ij} , \nonumber
\end{eqnarray}
where we have defined
\begin{equation}
[\beta]={{\hat R_y(\beta)}\over {\left<b\right| \hat R_y(\beta)
\left|a\right>}}, \nonumber
\end{equation}
and
\begin{equation}
\left<b\right|\hat R_y(\beta)\left|a\right> = [{\rm det}~
U(\beta)]^{1/2}.
\label{DET}
\end{equation}
Eqs. (\ref{ABCmatrix}) and (\ref{DET}) are written in a compact
form of $N\times N$ matrix, with $N$ being the number of total
single particles. The general principle of finding $U(\beta)$ and
$V(\beta)$ is given by the Thouless theorem \cite{Thouless60}, and
a well worked-out scheme can be found in the work of Tanabe {\it
et al.} \cite{Tanabe99}.

To write the matrices $U(\beta)$ and $V(\beta)$ explicitly, we
consider the fact that $\{a_i, a^\dagger_i\}$ and $\{b_i,
b^\dagger_i\}$ can both be expressed by the spherical
representation $\{c_i,c^\dagger_i\}$ through the HFB
transformation
\begin{equation}
\begin{array}{rcl}
\left[ \begin{array}{c} c \\ c^\dagger
\end{array} \right]
&=& \left( \begin{array}{cc} U_a & V_a \\ V_a & U_a
\end{array} \right)
\left[ \begin{array}{c} a \\ a^\dagger
\end{array} \right]
\vspace{8pt}\\
\left[ \begin{array}{c} c \\ c^\dagger
\end{array} \right]
&=& \left( \begin{array}{cc} U_b & V_b \\ V_b & U_b
\end{array} \right)
\left[ \begin{array}{c} b \\ b^\dagger
\end{array} \right].
\label{abcd}
\end{array}
\end{equation}
$U_a, V_a, U_b$ and $V_b$ in above equations, which define the HFB
transformation, are obtained from the Nilsson-BCS calculation. A
rotation of the spherical basis can be written in a matrix form as
\begin{eqnarray}
\hat R_y(\beta) \left[ \begin{array}{c} c \\ c^\dagger
\end{array} \right]
\hat R_y^\dagger (\beta) &=& \left( \begin{array}{cc} d(\beta) & 0
\\ 0 & d(\beta)
\end{array} \right)
\left[ \begin{array}{c} c \\ c^\dagger
\end{array} \right].
\label{transf}
\end{eqnarray}
Combining Eqs. (\ref{abcd}) and (\ref{transf}) and noting the
unitarity of the HFB transformation, one obtains
\begin{widetext}
\begin{eqnarray}
\hat R_y(\beta) \left[ \begin{array}{c} b \\ b^\dagger
\end{array} \right]
\hat R_y^\dagger (\beta) &=& \left( \begin{array}{cc} U_b & V_b \\
V_b & U_b
\end{array} \right)^T
\left( \begin{array}{cc} d(\beta) & 0 \\ 0 & d(\beta)
\end{array} \right)
\left( \begin{array}{cc} U_a & V_a \\ V_a & U_a
\end{array} \right)
\left[ \begin{array}{c} a \\ a^\dagger
\end{array} \right].
\end{eqnarray}
$U(\beta)$ and $V(\beta)$ can finally be obtained from the
following equation
\begin{equation}
\begin{array}{rcl}
\left( \begin{array}{cc} U(\beta)&V(\beta)\\V(\beta)&U(\beta)
\end{array} \right)
&=& \left( \begin{array}{cc} U^T_b & V^T_b \\ V^T_b & U^T_b
\end{array} \right)
\left( \begin{array}{cc} d(\beta) & 0 \\ 0 & d(\beta)
\end{array} \right)
\left( \begin{array}{cc} U_a & V_a \\ V_a & U_a
\end{array} \right)
\vspace{10pt}\\
&=& \left( \begin{array}{cc} U^T_bd(\beta)U_a+V^T_bd(\beta)V_a &
U^T_bd(\beta)V_a+V^T_bd(\beta)U_a
\vspace{6pt}\\
V^T_bd(\beta)U_a+U^T_bd(\beta)V_a &
U^T_bd(\beta)U_a+V^T_bd(\beta)V_a
\end{array} \right).
\end{array}
\end{equation}
\end{widetext}

We identify the following symmetry property for the matrix element
(\ref{RMT})
\begin{eqnarray}
&& \langle\Phi_{\kappa'}|\hat \beta^{\pm}_{1\mu}\hat
R(\beta)|\Phi_{\kappa}\rangle = (-)^{K_\kappa-K'_{\kappa'}}
\nonumber\\ &&\times \sum_{\mu'}d^1_{\mu'-\mu}(\beta)
\langle\Phi_{\kappa}|\hat \beta^{\mp}_{1\mu'}\hat
R(\beta)|\Phi_{\kappa'}\rangle .
\end{eqnarray}
In addition, there is a symmetry relation for the reduced matrix
element
\begin{eqnarray}
\langle \Psi_{I_f}\|\hat \beta^{\pm}\|\Psi_{I_i}\rangle =
(-)^{I_f-I_i}\sqrt{\frac{2I_i+1}{2I_f+1}}\langle \Psi_{I_i}\|\hat
\beta^{\mp}\|\Psi_{I_f}\rangle.
\end{eqnarray}
These symmetry properties can be used for testing the coding and
checking the numerical accuracy because the matrix elements at
each side of the equations are calculated independently.

\section{Gamow-Teller transition rates in a rare-earth example}

The formalism derived in the preceding section is valid for
general $\beta$-decay and electron-capture calculations no matter
whether a process is associated with an allowed or a forbidden
transition. Our following illustrative example considers allowed
transition only, for which the following selection rules (change
in quantum numbers between the initial and the final state) apply
\begin{equation}
\begin{array}{rcl}
\Delta I&=&0,\pm 1 , \vspace{6pt}\\
\Delta\pi&=&+1 .\nonumber
\end{array}
\end{equation}
We take the odd-odd nucleus $^{164}$Ho$_{97}$ as the example,
which decays to $^{164}$Dy$_{98}$ through an electron-capture
process by changing a proton to a neutron. We shall evaluate B(GT)
values defined in Eq. (\ref{BGT}). For the $A=164$ nuclei, the GT
interaction strengths in Eq. (\ref{GTstrength}) become
\begin{equation}
\begin{array}{rcl}
\chi_{GT}&=&{23 / A} = 0.14 ~ \mbox{(MeV)}
\vspace{4pt}\\
\kappa_{GT}&=&{7.5 / A} = 0.046 ~ \mbox{(MeV)}.
\end{array}\nonumber
\end{equation}
In order to compensate for the dependence of the decay rate on the
transition energy, it is customary to express the transition
probability in terms of the product $ft$ \cite{BMbook},
\begin{eqnarray}
ft=\frac{6163.4}{\left({g_A\over g_V}\right)^2_{\mbox{eff}}
B(GT)}, \label{ft}
\end{eqnarray}
where $t$ is the half-life and $f$ is a dimensionless quantity
depending on the charge of the nucleus and the energy and
multipolarity of the transition. For the coupling constants
appearing in Eq. (\ref{ft}), we simply adopt those from Ref.
\cite{Const} (see also the references cited therein),
\begin{eqnarray}
\left({g_A\over g_V}\right)_{\mbox{eff}}=0.74 \ {g_A\over g_V},
~~~~~~~~{g_A\over g_V}=-1.26 .\nonumber
\end{eqnarray}

\subsection{Structure description: Energy levels and electromagnetic transitions}

We use the standard Nilsson scheme \cite{Nilsson69} to generate
deformed single particle states for our basis. For the Nilsson
parameters we use those of Jain {\it et al.} \cite{Jain90}, which
are applicable to the rare earth region. The deformed bases are
obtained with the quadrupole deformation $\varepsilon_2=0.28$. At
this deformation, the Fermi levels of the $A=$ 164 nuclei are
surrounded by several neutron and proton single particle orbitals,
such as $\nu[523]{5\over 2}$ and $\pi[523]{7\over 2}$, which are
the most relevant ones in our later discussion on GT-transition.
The monopole-pairing coupling constants in Eq. (\ref{Mpairing})
are taken to be $G_1=20.12$ MeV and $G_2=13.13$ MeV, which are the
same values employed in many previous rare earth calculations. The
quadrupole-pairing strength $G_Q$ is proportional to $G_M$, the
proportionality constant being fixed to be 0.20 for the nuclei
considered in this paper.

\epsfigbox{fg 1}{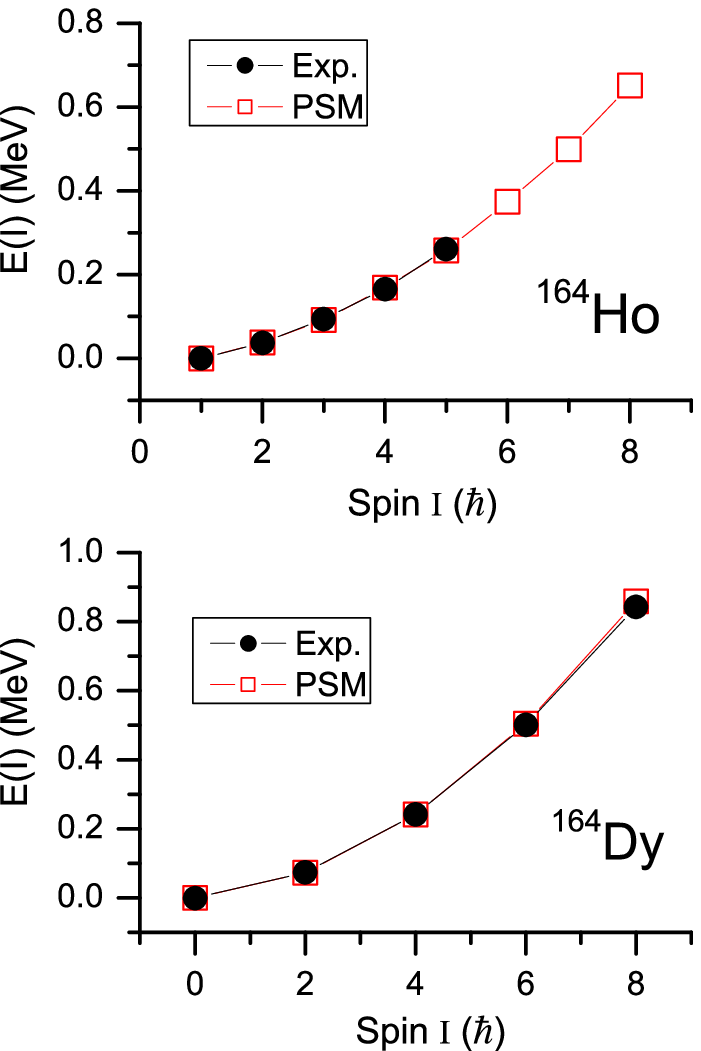}{0pt}{0pt} {(Color online) Calculated
energy levels of the ground band in $^{164}$Ho (upper panel) and
$^{164}$Dy (lower panel) are compared with data. The ground state
of the odd-odd $^{164}$Ho nucleus has $I^\pi = 1^+$.}

\epsfigbox{fg 2}{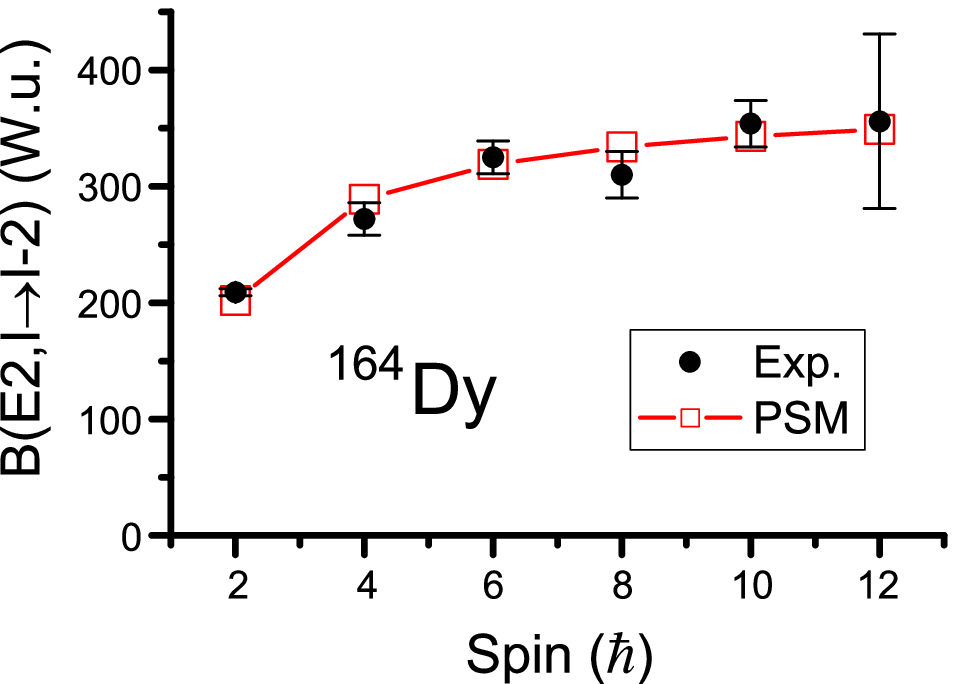}{0pt}{0pt} {(Color online) Calculated
B(E2) values of the ground band in $^{164}$Dy are compared with
data.}

Under the above calculation conditions, energy levels and wave
functions are obtained for the parent $^{164}$Ho and the daughter
$^{164}$Dy. In Fig. 1, we compare the calculated energy levels of
the ground bands with data. An excellent agreement between theory
and experiment is achieved, implying that with this established
set of parameters, the structure of these nuclei can be
microscopically described. In Fig. 2, one sees that the calculated
E2 transition probabilities along the ground band in $^{164}$Dy
also agree well with data. These B(E2) values are calculated with
the resulting wave functions as
\begin{equation}
B(E2, I_i\rightarrow I_f) = {{2I_f+1}\over{2I_i+1}} \left|\langle
\Psi_{I_f} || \hat Q_2 || \Psi_{I_i}\rangle\right|^2,
\end{equation}
where the effective charges 1.5$e$ for protons and 0.5$e$ for
neutrons are used.

\subsection{The Ikeda sum-rule}

The above calculations as well as those of the previous PSM
publications indicate that for low-lying states in heavy, deformed
nuclei, where one has a lot of structure data to compare with, the
PSM can usually provide a good description. For GT transitions, a
complete description requires the model to cover those highly
excited states as well because the GT strength usually spreads for
a wide range of excitation and a large amount of GT strength can
concentrate in some high energy states. Next, we examine the
resulting wave functions of both low-lying and high-lying states
in the full model space. There is a well-known charge-exchange
sum-rule, the Ikeda sum-rule \cite{Ikeda65}, expressed as
\begin{eqnarray}
&&S(\mathrm{GT^-})-S(\mathrm{GT^+})\nonumber\\
&=&\sum_f B(\mathrm{GT^-},i\rightarrow f)-\sum_f B(\mathrm{GT^+},i\rightarrow f)\nonumber\\
&=&\sum_{f,\mu}|\langle\Psi_f|\hat \beta^-_{1\mu}|\Psi_i\rangle|^2
-\sum_{f,\mu}|\langle\Psi_f|\hat \beta^+_{1\mu}|\Psi_i\rangle|^2\nonumber\\
&=&3(N-Z).
\label{Ikeda}
\end{eqnarray}
This model independent sum-rule results from the commutation
relation between the isospin operators. Hence, a
(non-relativistic) model is expected to satisfy it as long as the
basis in Eq. (\ref{Ikeda}) is complete.

\epsfigdox{fg 3}{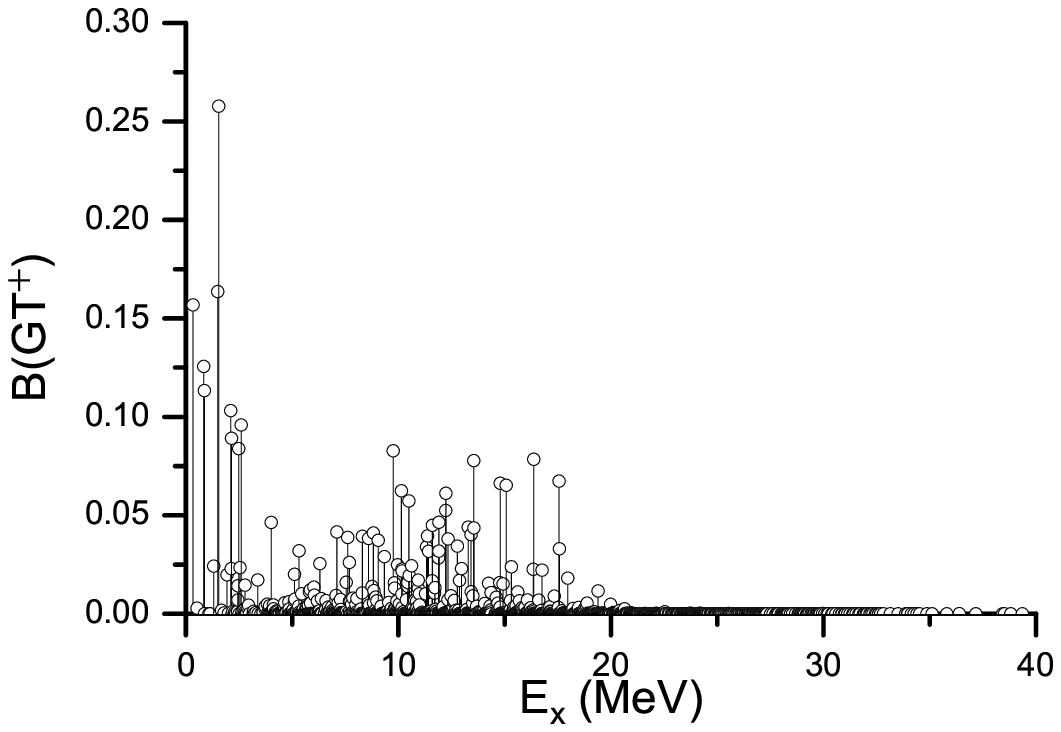}{0pt}{0pt} {Distribution of the
B(GT$^+$) strength.}

\epsfigdox{fg 4}{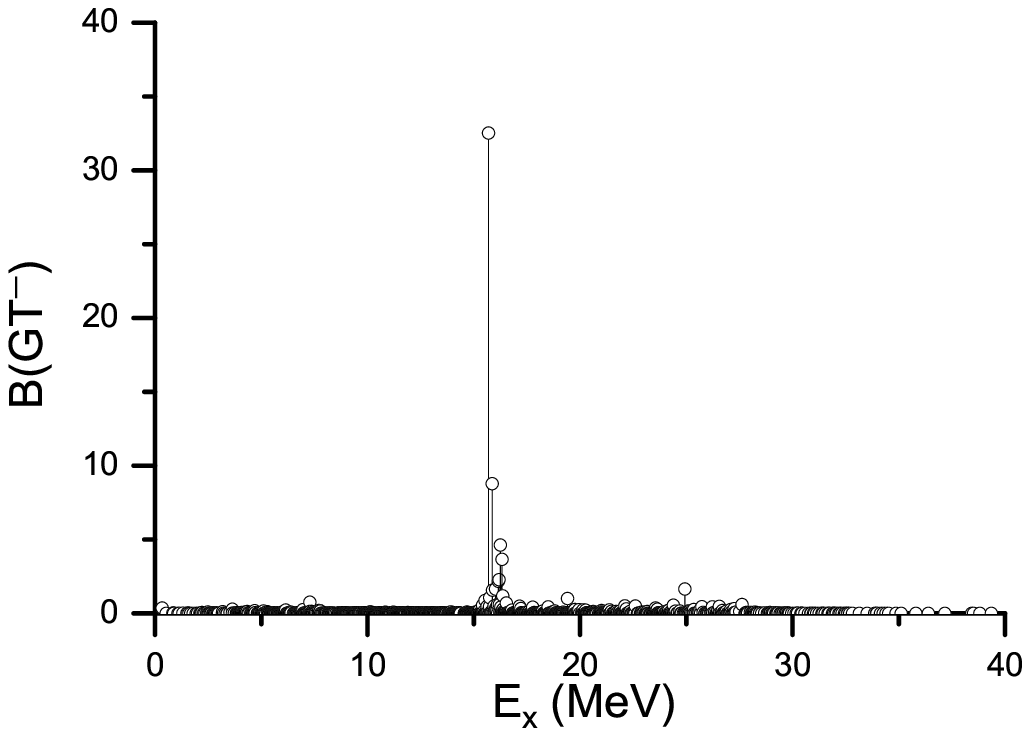}{0pt}{0pt} {Distribution of the
B(GT$^-$) strength.}

\epsfigdox{fg 5}{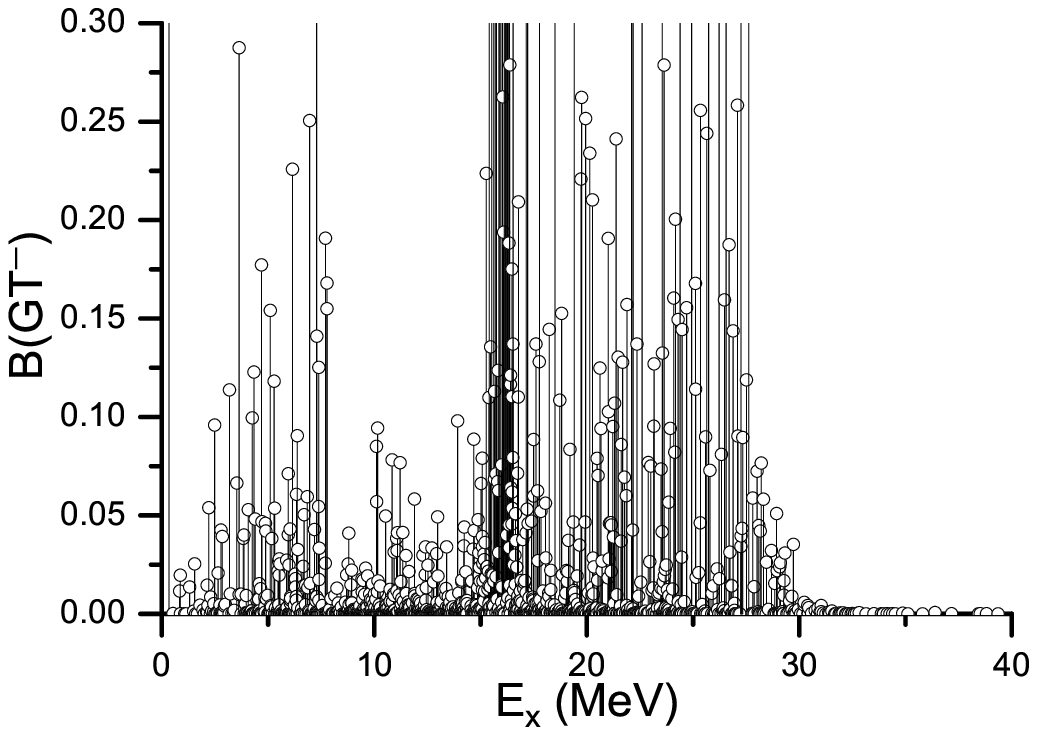}{0pt}{0pt} {Distribution of the
B(GT$^-$) strength. The numbers are the same as those of Fig. 4,
but plotted in the scale as Fig. 3.}

Let $|\Psi_i\rangle$ in Eq. (\ref{Ikeda}) correspond to the 0-qp
ground state of a nucleus and $|\Psi_f\rangle$ to the $f$-th 2-qp
state with $I^\pi=1^+$. The summation in Eq. (\ref{Ikeda}) runs
over all possible 2-qp $1^+$ states in the complete model space.
When the single-particle states with $N=$ 3, 4, 5, 6
harmonic-oscillator shells are taken into account for both
neutrons and protons, the total number of $1^+$ states turns out
to be 1718. For our example of $^{164}$Dy ($N=98$ and $Z=66$), we
show in Figs. 3, 4, and 5 all the obtained B(GT$^+$) and B(GT$^-$)
strengths as functions of the $1^+$ state energy. In Fig. 3, it is
observed that a considerable portion of the B(GT$^+$) strength
appears at the low excitation region less than 2.5 MeV. The rest
of the strength is fragmented over many states extended to about
20 MeV. The fragmentation is clearly caused by the residual
interactions. In Fig. 4, a strong peak of the B(GT$^-$) strength
appears around 16 MeV, while the rest of the strength spreads for
nearly the entire energy region. The concentration around 16 MeV
takes most of the B(GT$^-$) strength, and has a character of GT
giant resonance. In Fig. 5, we show the same B(GT$^-$)
distribution as that of Fig. 4, but with a different scale to
emphasize those smaller strengths. It is seen that the
fragmentation extends to the excitation of 30 MeV.

Adding respectively all the B(GT$^+$) and B(GT$^-$) strengths
together, as indicated in Eq. (\ref{Ikeda}), we get
\begin{eqnarray}
S(\mathrm{GT^-})=97.4957, ~~~~ S(\mathrm{GT^+})=4.1717. \nonumber
\end{eqnarray}
This means that our $S(\mathrm{GT^-})-S(\mathrm{GT^+})$ exhausts
97.2$\%$ of the sum-rule value $3(N-Z)$. Note that in our Nilsson
calculation for deformed single particle states, we use the
standard formalism \cite{Nilsson69} in which the neutron
harmonic-oscillator frequency $\omega_\nu$ differs from the proton
frequency $\omega_\pi$ for $N\ne Z$ nuclei. If we set
$\omega_\nu=\omega_\pi$, the calculated values become
\begin{eqnarray}
S(\mathrm{GT^-})=97.7118, ~~~~ S(\mathrm{GT^+})=1.7797, \nonumber
\end{eqnarray}
which approaches 99.9$\%$ of the sum-rule value. If we further
switch off all the pairing interactions, i.e. if we abandon the
quasiparticle picture and go back to the particle picture, we get
\begin{eqnarray}
S(\mathrm{GT^-})=96.7480, ~~~~ S(\mathrm{GT^+})=0.7480. \nonumber
\end{eqnarray}
Thus, in the idealized (but not realistic) case in which neutrons
and protons move in the same harmonic-oscillator potential well
with no pairing interaction, the Ikeda sum-rule (\ref{Ikeda}) is
fulfilled exactly. The above results may be taken as an additional
nontrivial check for the coding since in all these calculations,
the wave functions in Eq. (\ref{Ikeda}) have a form of Eq.
(\ref{wf}) and the B(GT) values are computed in a sophisticated
way as presented in Sections II and III.

\subsection{Effects of the Gamow-Teller forces and
configuration mixing}

The GT interaction in Eq. (\ref{GT}) is a new addition to the PSM
Hamiltonian (\ref{Hamilt}). The results in Figs. 1 and 2 seem to
indicate that the GT forces do not have effects on the low-lying
states since with or without the GT forces, the energy levels and
electromagnetic transitions are described equally well. The
purpose of this section is to study how the new addition of the GT
forces influences the B(GT) values and in what extent the
configuration mixing modifies the results.

\begin{table}
\caption{Five different cases of the GT transition calculation. In
the table, ``even-even" means the $0^+$ ground state of $^{164}$Dy
and ``odd-odd" means the $1^+$ ground state of $^{164}$Ho.
$H_{GT}$ stands for the Gamow-Teller forces in the Hamiltonian.}
\label{tab:1}
\begin{tabular}{c|rl|rl|c}
\hline\noalign{\smallskip}
Case    & Even-even  && Odd-odd   && $H_{GT}$ \\
\noalign{\smallskip}\hline\noalign{\smallskip}
1 & $\{|0\rangle\}$ & (1) & $\{ \nu[523]{5\over 2}\times \pi[523]{7\over 2}\}$ & (1) & no \\
2 & $\{|0\rangle, |2\nu 2\pi\rangle\}$ & (857) & $\{|1\nu 1\pi\rangle\}$ & (823) & no \\
3 & $\{|0\rangle\}$ & (1) & $\{|1\nu 1\pi\rangle\}$ & (823) & yes \\
4 & $\{|0\rangle, |2\nu 2\pi\rangle\}$ & (322) & $\{|1\nu 1\pi\rangle\}$ & (823) & yes \\
5 & $\{|0\rangle, |2\nu 2\pi\rangle\}$ & (857) & $\{|1\nu
1\pi\rangle\}$ & (823) & yes \\
\noalign{\smallskip}\hline
\end{tabular}
\end{table}

\epsfigpox{fg 6}{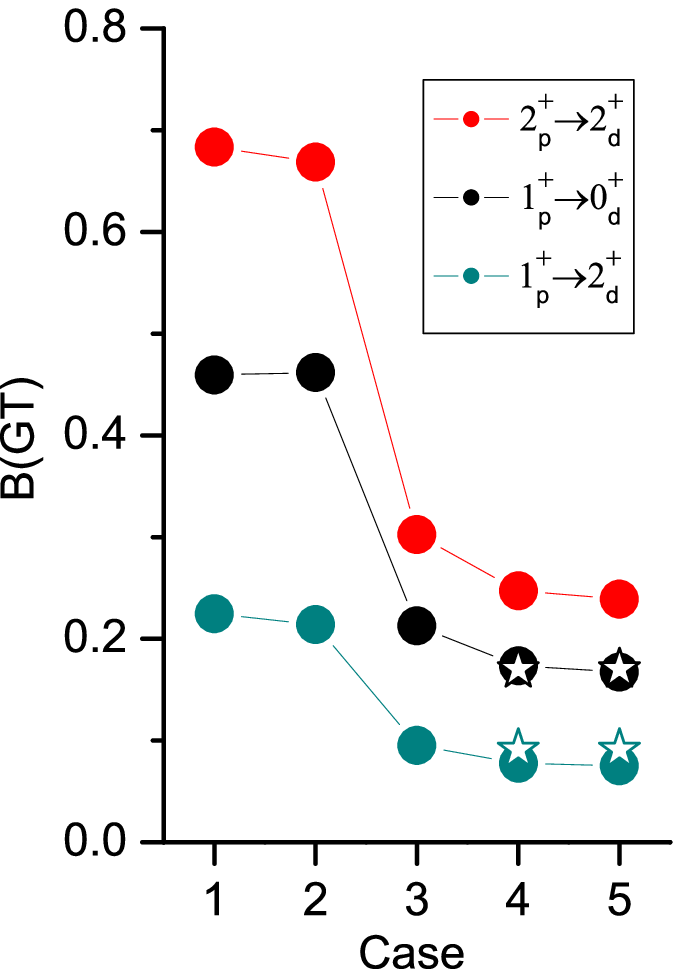}{0pt}{0pt} {(Color online) B(GT) values
calculated for the five different cases. Available data are shown
in stars. The B(GT) data are converted from the experimental
Log$ft$ values \protect\cite{Data} through Eq. (\ref{ft}).}

We perform calculations for the GT transition from the $1^+$
ground state of $^{164}$Ho to the $0^+$ ground state of $^{164}$Dy
with five different cases as listed in Table I, and results are
shown in Fig. 6. For Cases 1 and 2, we do not include the GT
forces. In Case 1, there is only one dimension in both the parent
$^{164}$Ho and the daughter $^{164}$Dy wave function; namely, we
approach the wave function of the even-even nucleus using the
projected vacuum state and that of the odd-odd nucleus using the
projected 2-qp state $\{ \nu[523]{5\over 2}\times \pi[523]{7\over
2}\}$. In Case 2, we include 823 2-qp basis states in the odd-odd
and additional 856 4-qp states in the even-even wave function. As
one can see from Fig. 6, there is no much difference between Case
1 and Case 2, implying that without the GT forces, configuration
mixing becomes unimportant. It is also seen in these two cases
that without the GT forces, the calculated B(GT) is by nearly a
factor of three too large when compared with data.

A significant reduction in B(GT) is obtained when we switch on the
GT forces. As shown in Fig. 6 for Cases 3, 4, and 5, these forces
push down the B(GT) values considerably. Moreover, effect of
configuration mixing shows up when the GT forces are included;
there is a clear difference between the results of Case 3 (with no
4-qp state in the even-even wave function) and Case 4 (with 321
4-qp basis states mixed in the even-even wave function). The
latter case reproduces data well. Comparing the result of Case 5
with Case 4, one sees a converging trend in the calculation;
further enlarging the configuration space would no longer improve
the result as far as the GT transition between the ground states
is concerned.

\begin{table}
\caption{Comparison of calculated Log$ft$ values with data for
electron-capture from $^{164}$Ho to $^{164}$Dy and $\beta^-$ decay
from $^{164}$Ho to $^{164}$Er. Data are taken from Ref.
\protect\cite{Data}.} \label{tab:2}
\begin{tabular}{c|cc}
\hline\noalign{\smallskip}
Decay  & Log$ft$ (Th.) & Log$ft$ (Exp.) \\
\noalign{\smallskip}\hline\noalign{\smallskip}
$^{164}$Ho ($1^+$) $\rightarrow$ $^{164}$Dy ($0^+$) & 4.61 & 4.59 (8) \\
$^{164}$Ho ($1^+$) $\rightarrow$ $^{164}$Dy ($2^+$) & 4.96 & 4.86 (6) \\
\noalign{\smallskip}\hline\noalign{\smallskip}
$^{164}$Ho ($1^+$) $\rightarrow$ $^{164}$Er ($0^+$) & 5.19 & 5.54 (10)\\
$^{164}$Ho ($1^+$) $\rightarrow$ $^{164}$Er ($2^+$) & 5.62 & 5.75 (10)\\
\noalign{\smallskip}\hline
\end{tabular}
\end{table}

$^{164}$Ho can capture an electron and change to $^{164}$Dy, and
can $\beta^-$-decay to $^{164}$Er. In Table II, we compare
calculated Log$ft$ values with the existing data for both
processes. We have used the same parameters in generating the
$^{164}$Er wave functions and in the calculation of the transition
matrix elements. As can be seen from Table II, the agreement in
both processes is satisfactory.

\subsection{GT transition rates of the excited states}

\epsfigbox{fg 7}{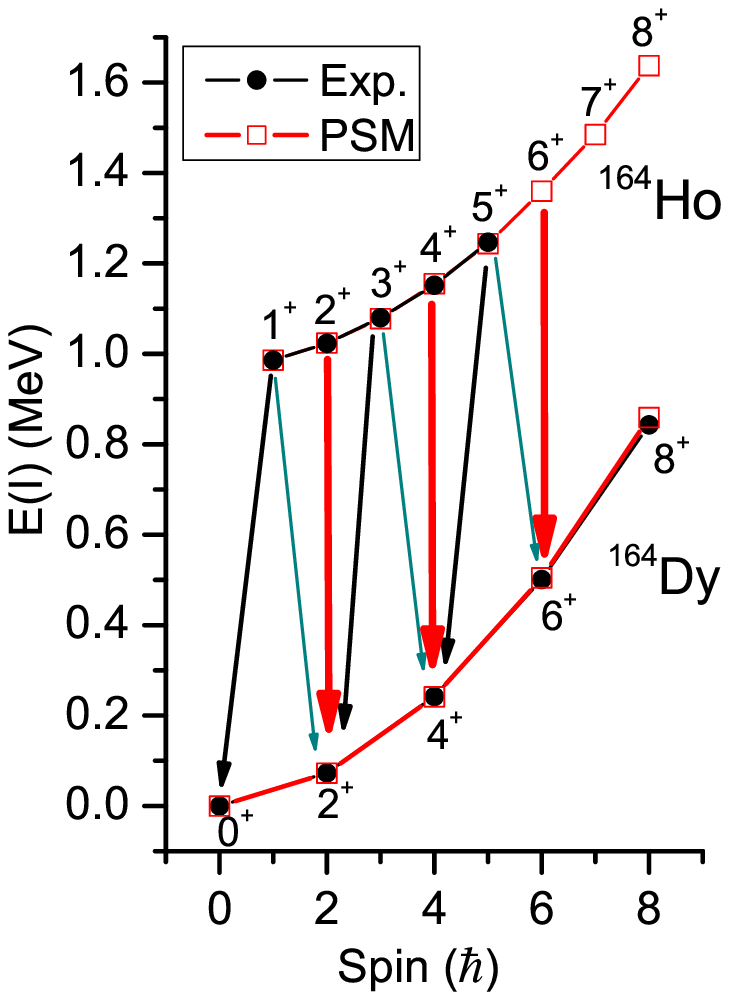}{0pt}{0pt} {(Color online) Possible
transition paths from the ground band of $^{164}$Ho to that of
$^{164}$Dy. Experimentally known energies for these states are
also displayed.}

In Fig. 7, possible allowed GT transitions from the ground band of
$^{164}$Ho to that of $^{164}$Dy are illustrated. Here, energy
levels up to about 800 keV of excitation in both parent and
daughter nuclei are considered. They are $\Delta I=I_p - I_d = +1$
transitions (in black), $\Delta I = 0$ transitions (in red), and
$\Delta I = -1$ transitions (in green). The calculated B(GT)
values are shown in the upper panel, and the corresponding Log$ft$
values in the lower panel of Fig. 8. The experimentally measured
decay probabilities are those of the $I_p = 1 \rightarrow I_d = 0$
and $I_p = 1 \rightarrow I_d = 2$ transitions, with which our
calculation agrees well. The other transition probabilities
associated with the excited states are our prediction. Note that
the decay rates with $\Delta I = 0$ (in red) are predicted to have
larger B(GT) and smaller Log$ft$ values than the measured $\Delta
I=\pm 1$ transitions. This is an interesting result because
normally, it is very difficult to study $\beta$-decay rates of
excited states in laboratory since excited nuclear states decay
much faster via $\gamma$-emission than by $\beta$-decay. A
measurement may be possible only if the underlying states are
long-lived isomers.

\epsfigbox{fg 8}{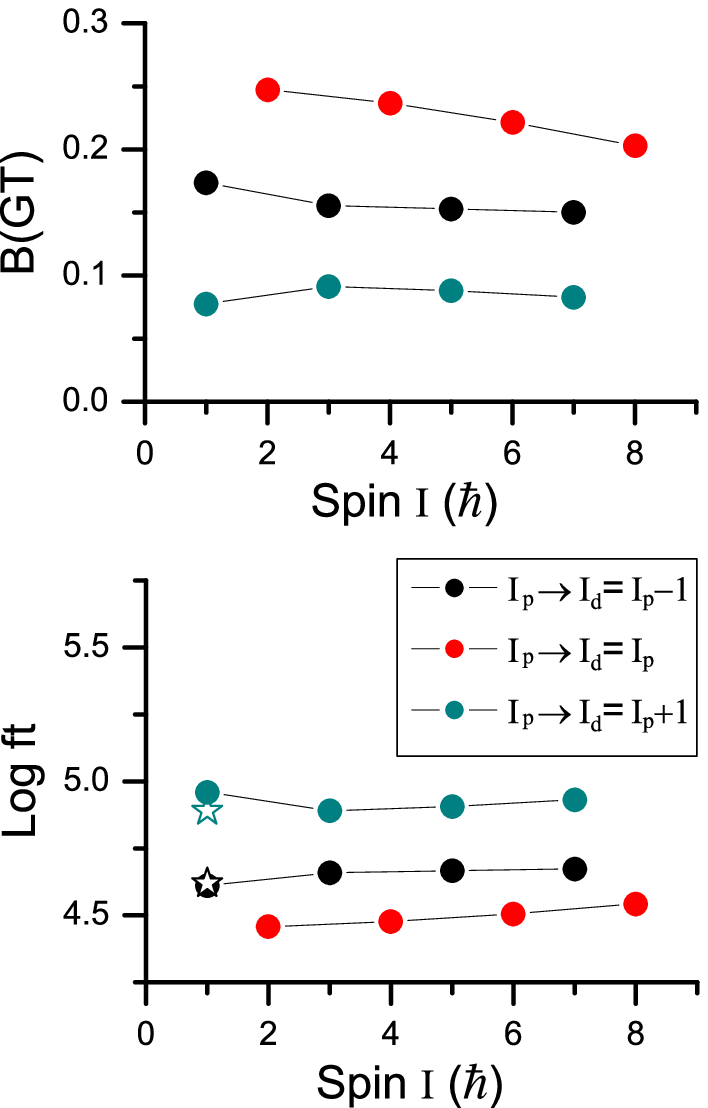}{0pt}{0pt} {(Color online) Calculated
B(GT) (upper panel) and Log$ft$ values (lower panel) for the
$^{164}$Ho $\rightarrow$ $^{164}$Dy electron-capture process.
Available data are shown in stars. Data are taken from Ref.
\protect\cite{Data}.}

In stellar environments, nuclei that are exposed to high
temperatures and high densities may experience a significant
enhancement in their decay rate. This idea was brought out early
by Cameron \cite{Cameron59} and studied in detail by Bahcall
\cite{Bahcall61}. The enhancement can result from the effects
related to the thermal population of excited nuclear states in the
hot photon bath. In a thermal equilibrium with temperature $T$,
the population probability is determined by the Boltzmann factor
and the statistical weight
\begin{equation}
p_i = {{(2I_i+1)e^{-E_i/kT}}\over {\sum_m(2I_m+1)e^{-E_m/kT}}}.
\end{equation}
Hence the actual stellar decay rate must consider decays of all
thermally populated excited states $i$ in the parent nucleus to
the accessible levels $j$ of the daughter. Thermal enhancement of
beta-decay rates can have substantial impact on nucleosynthesis,
for example, the s-process branching conditions
\cite{Kaeppeler99}.

\section{Summary and outlook}

In this article, we have presented the new development of a shell
model method for calculation of Gamow-Teller transition rates. The
method is based on the Projected Shell Model. Different from the
conventional shell model, which builds its states in a spherical
basis, the PSM constructs its states in a deformed basis in which
important nuclear correlations are taken into account very
efficiently. Therefore, it is possible for a shell model
diagonalization in the PSM to be carried out in a manageable space
for medium to heavy, and even for super-heavy nuclei.

We have shown how the GT transition matrix elements are calculated
in the PSM framework. A computer code has been developed and been
tested. One nontrivial test has been done through the Ikeda
sum-rule. We have obtained a reasonable distribution of the B(GT)
strength with a fulfillment of the sum-rule. We have presented the
first example from the rare earth region. In the calculation of
$^{164}$Ho $\rightarrow$ $^{164}$Dy electron-capture process, we
have predicted the GT transition rates for the excited states.
Such rates should be included as part of the total rate when these
states are thermally populated in hot stellar environments.

In the calculation discussed so far, we have considered only
allowed $\beta$-decay for the low-lying states. Calculation of
allowed $\beta$-decay for states with high excitation and the
forbidden transitions is possible in the PSM framework. Study of
GT giant resonance is also under consideration.

The method described in the present article can be applied to
various fields such as nuclear astrophysics and fundamental
physics, where weak interaction processes take place in nuclear
systems \cite{ALW05}. In particular, one may find interesting
applications to cases where a laboratory measurement for certain
weak interaction rates is difficult and where the conventional
shell model calculations are not feasible. Potential applications
in nuclear astrophysics are calculations of $\beta$-decay rates
for the r-process \cite{r-process} and the rp-process
\cite{rp-process} nucleosynthesis, and electron-capture rates for
the core collapse supernova modelling \cite{LM03,SN}. In the
double-$\beta$ decay theory, theoretical calculations for the
nuclear matrix elements are needed, for which one has relied on
the Quasiparticle Random Phase Approximation
\cite{Civitarese98,double-beta}, particularly when heavy nuclei
are involved. We expect that the method presented here can make
important contributions to all these studies.

\section{Acknowledgement}

The authors are grateful to the Joint Institute for Nuclear
Astrophysics (JINA) for support. Z.-C. G. thanks H. Schatz and M.
Wiescher for the warm hospitality. Y.S. thanks J. Hirsch for many
suggestive discussions. Comments and suggestions from A. Arima, B.
A. Brown, J. Engel, and V. Zelevinsky are acknowledged. This work
is partly supported by NNSF of China under contract No. 10305019,
10475115, 10435010, by MSBRDP of China (G20000774), and by NSF of
USA under contract PHY-0140324 and PHY-0216783.

\baselineskip = 14pt
\bibliographystyle{unsrt}

\end{document}